# Estudio arqueoastronómico de los templos andinos de Arica y Parinacota, Chile

Alejandro Gangui (Argentina), Ángel Guillén (Chile) y Magdalena Pereira (Chile)

El estudio de la orientación de iglesias, junto con el de las pirámides de Egipto y los monumentos megalíticos europeos, se encuentra entre los temas más antiguos y más trabajados de la arqueoastronomía. Se sabe que la orientación espacial de las iglesias cristianas antiguas es una de las características más salientes de su arquitectura. En el continente europeo y en multitud de sitios lejanos a donde llegaron los evangelizadores, existe una marcada tendencia a orientar los altares de los templos en el rango solar. Es decir, el eje del templo, desde la puerta principal y en dirección al altar, se orienta hacia aquellos puntos del horizonte por donde sale el Sol en diferentes días del año. Entre estos días, existe una notoria preferencia por aquellos correspondientes a los equinoccios astronómicos, cuando las orientaciones son cercanas al este geográfico (McCluskey 1998). Sin embargo, dentro del mismo rango solar no son infrecuentes las alineaciones en sentido contrario, con el altar a poniente (Esteban et al. 2001; Belmonte et al. 2007), aunque resultan excepcionales pues no siguen el patrón canónico.

Pasado un cierto tiempo del asentamiento español en el virreinato del Perú, el virrey Francisco de Toledo decide reorganizar el territorio y su población (Toledo 1986). Dedica una atención especial a las rutas comerciales del ámbito surandino, cuyo principal objetivo era el de procurar la salida de la plata desde Potosí al Pacífico y el transporte del azogue desde Huancavelica a las minas alto-andinas. Fue así que, camino hacia Arica, puerto oficial de salida de la plata y entrada del azogue desde 1574, se formaron pequeños caseríos y tambos con poblaciones estables. Las iglesias andinas de esta región surgieron en sitios estratégicos a lo largo de la ruta que trajinantes recorrían para transportar los metales preciosos desde Potosí hacia las playas de Arica, en particular en torno a los valles de Lluta y Azapa. Al ser una zona extensa y difícil de transitar, la influencia de los doctrineros en la configuración de los poblados y reducciones indígenas fue limitada. Este hecho acentuó el diálogo entre las culturas locales y la tradición occidental, y esta interacción pudo ser el origen de muchas características propias de la arquitectura andina; las iglesias construidas en épocas coloniales forman un corpus de estudio importante para intentar develar algunos elementos de dicha interacción.

Hemos estudiado un conjunto grande de iglesias patrimoniales de la región, concentrando nuestra atención en sus orientaciones, emplazamientos geográficos y en el paisaje que las rodea (Fundación Altiplano 2012). Medimos la orientación de un total de 38 iglesias andinas, la mayoría de ellas reconstruidas durante el siglo XIX, ubicadas en el altiplano, sierra y valles bajos de las quebradas de Lluta, Azapa, Vítor y Camarones (Fig. 1). Como veremos, la muestra indica que, aunque no se siguió un patrón único de orientación determinante en toda la región, casi la mitad de las iglesias estudiadas se orienta dentro del rango solar, con una proporción dominante en aquellas que presentan su altar hacia el poniente.

En otros trabajos nos hemos concentrado en el contexto histórico y cultural de la región estudiada y en el rol importante que en ésta ha desempeñado la ruta de la plata (Moreno y Pereira 2011; Gangui, Guillén y Pereira 2016). En las próximas secciones hacemos una reseña breve sobre la república de indios, la religiosidad andina y sobre las iglesias de los Altos de Arica. En la última parte de nuestro trabajo detallamos los estudios arqueoastronómicos llevados a cabo en lo referente a las orientaciones de los templos y analizamos nuestros resultados preliminares. Concluimos con una breve discusión sobre los patrones de orientación encontrados en nuestras mediciones y sobre sus posibles causas.

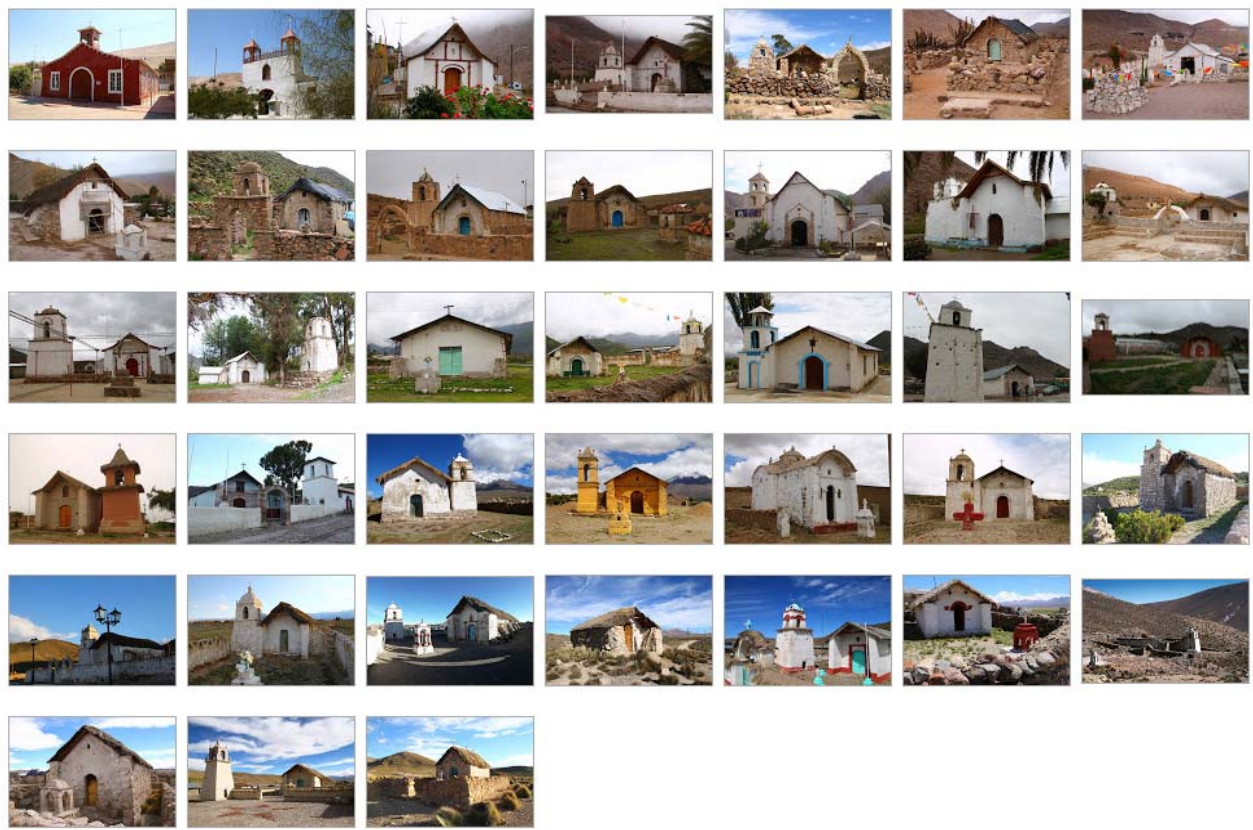

Figura 1. Las treinta y ocho iglesias patrimoniales medidas en este estudio. De arriba abajo y de izquierda a derecha, las fotografías corresponden a las iglesias de las siguientes localidades: Azapa, Poconchile, Chitita, Guañacagua, Sucuna, Saguara, Pachica, Esquiña, Aico, Parcohaylla, Mulluri, Codpa, Timar, Cobija, Timalchaca, Tignamar, Chapiquiña, Pachama, Saxamar, Belén (Santiago Apóstol y Virgen de la Candelaria), Socoroma, Putre, Tacora, Airo, Chapoco, Putani, Pucoyo, Cosapilla, Guacollo, Parinacota, Ungallire, Caquena, Chucuyo, Choquelimpie, Ancuta, Guallatire y Churiguaya.

**La república de indios: idea físico-espacial del territorio occidental sudamericano**

Un territorio impregnado de simbolismos es el que encontraron los colonizadores europeos al desembarcar en América. Nuevos patrones culturales, sociales y económicos fueron una referencia ineludible que exigió un arduo trabajo para facilitar el arraigo de aquellos grupos humanos al espacio geográfico del nuevo mundo (Galdós Rodríguez 1985).

Desde el año 1560 funcionarios reales enfrentaron un importante proceso de reformas del gobierno colonial del Perú, dejando como resultado documentos donde dieron cuenta sobre el pasado indiano y los procedimientos que debían emplearse para gobernar convenientemente a las comunidades indígenas. Tales personalidades fueron encarnadas en su momento por el gobernador Lope García de Castro y el ya mencionado virrey Toledo, quienes asumieron las reales indicaciones de Felipe II. Un primer paso para la demarcación territorial fue afrontado por el gobernador desde el año 1565, creándose los Corregimientos, como los de Arica y Lluta. Desde ese momento los funcionarios de la corona asumieron la administración territorial de la región. Luego entra en escena Toledo, quien fue el artífice de una sistemática reingeniería social que sacudió a las debilitadas estructuras organizacionales andinas. El virrey ordenó el traslado total de las poblaciones indígenas desde sus aldeas hacia nuevos emplazamientos donde se fundaron los así llamados "pueblos de indios"

En el virreinato del Perú la introducción paulatina de las Ordenanzas Toledanas significó el declive progresivo de la economía andina. La clara intencionalidad mercantilista de la medida basada en un uso mayoritario de la población para los fines de la mita minera justificó sus contenidos. La "república de indios" se conformó en contraste con la "república de españoles", una respuesta concreta en el plano físico-espacial que apeló a un modelo geométrico donde se apreciaba un mensaje implícito y diferente: orden y policía. Toledo validó su impronta al mencionar reiteradamente en sus escritos el nivel de abandono y exaltación de vicios en que, supuestamente, se desenvolvía la población indígena. Una clara visión política y, desde luego, conveniente a los efectos de la implementación reduccional.

La participación de la iglesia, afianzada por los preceptos tridentinos, fue el apoyo perfecto para las estimaciones toledanas: debía difundirse sostenidamente el pensamiento religioso occidental, cualquier otra forma de espiritualidad predominante resultaba inoportuna. En contraste con el pensamiento dual andino, la visión occidental referida en un nuevo orden para las tierras americanas se enmarcaba en la concepción y plasmación de un sistema político integral, donde la uniformidad doctrinal debía prevalecer. El despliegue evangelizador y extirpador de idolatrías fue concurrente con una utilización de la propagandística religiosa para desactivar los arraigados simbolismos territoriales andinos y sirvió, paralelamente, para los objetivos de la corona en los aspectos culturales, económicos, políticos y sociales.

El envolvente urbanístico de la cuadrícula española se convirtió en el modelo idóneo para implementar las medidas de inclusión-exclusión. Los promotores de las disposiciones reglamentarias pretendían el vaciamiento aparente del territorio y la reconcentración, lo que significó obtener un catastro en tiempo real de las poblaciones y, además, asegurarse la disposición inmediata de tributarios para participar en los obrajes y mitas mineras. Los grupos humanos indígenas adquirieron un valor económico de orden estratégico y, en consecuencia, se impuso un renovado orden social con un espacio geográfico tensionado por una república de españoles (la ciudad) y una república de indios (los pueblos reduccionales). Así, esta dualidad explicitó la posición ventajosa de las ciudades y la situación marginal de los pueblos reducidos.

Una alianza estratégica para el logro de los objetivos reduccionales se dio con la actuación de los caciques, quienes valiéndose de la nueva estructura institucional desplegaron un proceso de auto colonización. Podemos preguntarnos si la estrategia reduccional alcanzó plenamente sus propósitos en el entorno andino. Creer que el desarraigo poblacional fue una operación inequívoca y definitiva de sumisión ideológica es una afirmación desacertada. La movilización de personas se convirtió en un acto aparente e insuficiente, subestimándose el grado de interiorización simbólica del ser andino con su territorio.

¿Qué aspectos procedentes de la impronta reduccional valdría contrastar en esta época? Ciertamente aquellos que todavía influyen en la vigencia y continuidad de las comunidades andinas, donde el tema dialéctico de la dispersión/encerramiento es tan destacable como la exégesis del espacio aplicado frente al espacio resignificado. Un acto pragmático de corte simbólico con implicancia urbanística fue el mantenimiento de la dualidad indígena plasmada en la vigencia de las parcialidades, que pese al afán desestructurador de las autoridades coloniales, contribuyeron eficazmente en la resignificación de los espacios, promoviendo la sacralización territorial de contenidos indígenas.

El modelo andino fue asombroso por la flexibilidad de su escala productiva. En un entorno geográfico tan complejo hubiera sido difícil subsistir con una producción absolutamente local. En contraste con esto, las sociedades andinas –pequeñas, medianas o grandes–, apostaron por una economía convergente dispuesta en "archipiélagos" que contribuyeron indistintamente con recursos marinos y/o terrestres. La mayor y mejor justificación para el florecimiento de las culturas locales estuvo en la comprensión de la dimensión ambiental del territorio.

El modelo geométrico aplicado para configurar a los nacientes pueblos indígenas respetó la usanza peninsular. El concepto político de la ciudad europea fue replicado y en cada pueblo se definió la existencia de dos alcaldes, cuatro regidores, un alguacil y un quipucamayo junto a otros funcionarios oficiales. Se propuso una periodicidad anual para la realización de procesos eleccionarios. Las personas que no habían aceptado el credo cristiano quedaron excluidas de la actuación municipal. Los pobladores que hubieran sido sancionados por razones eclesiásticas –como prácticas idolátricas– estuvieron imposibilitados para acceder a cargos públicos. La nueva política, definida desde España en la Junta Magna del año 1568 incidió en un nuevo orden, con el fin de incrementar la rentabilidad del virreinato peruano y mejorar la condición deficitaria de las arcas reales. Parte de esta visión gubernamental se afianzó en el control territorial que, apoyándose en la idea de orden y policía, estableció un reagrupamiento riguroso de las poblaciones indígenas, mejor conocido como reducciones toledanas (Málaga Medina 1989).

**Religiosidad andina y occidental**

En los años finales de la civilización andina, bajo la égida imperial de los incas, se mantuvieron con tolerancia la proliferación de los credos y prácticas rituales locales. El Sol y la Luna gozaron de prerrogativas especiales en el imaginario religioso andino. Con fines agrícolas y cotidianos la adoración a la Pachamama es, hasta la fecha, un acto devocional obligatorio y de profunda interiorización comunal. La prevalencia de un Dios (Wiracocha) autor del universo, es una referencia simbólica de identidad y divinidad celestial, caracterizadora de una imagen guerrera y revolucionaria, cuya proyección iconográfica fue empleada para conjugar los mensajes semióticos y semánticos cristianos y andinos.

Las ceremonias de advocación, para las divinidades indígenas, se realizaban con boato en grandes espacios y eran encaminadas por oficiantes, entregándose dádivas y ofrendas consistentes en comida e, incluso, oro y plata. Gran parte de esas salutaciones se reiteraban en el recuerdo permanente de las huacas y momias. La existencia de mecanismos de ayuno/penitencia como la confesión y/o simbólicos como la denominada "cruz andina" causó una serie de inquietudes en el pensamiento de los sacerdotes cristianos. En relación al mundo aymara, las conectividades religiosas territoriales fueron, en el caso surandino, convergentes hacia el lago Titicaca, particularmente con el actual Santuario de Copacabana, donde era vital aplacar, entre otras, las fuerzas del trueno (Llapta) y del rayo (Illapa) (Gisbert 1999). La bóveda celeste fue vital para los pueblos aymaras, sobre todo las constelaciones del sur que regían el decurso de la vida agrícola y cotidiana.

La monarquía española desde los Reyes Católicos, pasando por Carlos V y luego por su hijo y sucesor Felipe II, se apoyó para sus conquistas territoriales en los Derechos del Real Patronato o Patronato Regio, que implicaron un conjunto de bulas papales, denominadas también "Bulas de donación". Además, a Carlos V le correspondió apoyar la convocatoria del Concilio de Trento (1545-1563). Años más tarde, Felipe II, en el poder desde enero de 1556, afrontó conflictos de procedimiento con el papado y mantuvo discrepancias por la intensificación de las labores de la contrarreforma.

En el Concilio de Trento se estimó el aporte del arte sacro como un procedimiento de comunicación que debería incorporar posiciones simples y sencillas, argumentadas en la percepción e inclusión de los sentidos: era el preludio del barroco. Mientras tanto, desde el año 1551, en la ciudad de Lima empezaron los Concilios Limenses –cinco eventos desplegados hasta 1601– que encontraron su justificación en las exigencias de la nueva iglesia virreinal peruana, demandándose organización, jurisdicción, administración, arquitectura y arte. Los Concilios Limenses (o provinciales) fueron vitales para Felipe II, que en una carta del 9 de enero de 1566 dirigida a Toledo (quien tres años más tarde sería virrey del Perú) le expresaba: "Si la labor de Trento pareció siempre insuficiente a los españoles, se comprende cómo la reunión de los concilios provinciales les brindaba la oportunidad para aplicarla con celo. Porque en ello va toda la cristiandad, cuyo bien yo deseo y he de procurar siempre en primer lugar".

Aquellas directrices emanadas en el período tridentino para el desarrollo de las artes y la arquitectura religiosa tuvieron su contraparte desde los conclaves limeños. Así, se indicó que la erección de las fábricas tenía que realizarse en los lugares de concentración poblacional indígena. Durante el primer tercio del siglo XVII el sacerdote Vázquez de Espinosa viajó por el norte chileno y en el relato de su travesía reveló, además de significativos aspectos geográficos del borde occidental de América del Sur, el incipiente proceso de adoctrinamiento religioso que se experimentaba en los andes meridionales (Vázquez de Espinosa 1948 [1629]). La supuesta precariedad espiritual –que ya mencionamos sugería Toledo en sus escritos sobre la población indígena– tenía una contraparte física que evidenciaba la pervivencia de pueblos misérrimos con iglesias desvencijadas, inconclusas o sin edificar (Fig. 2). Vázquez de Espinosa mencionó también la existencia del Camino Real entre Arica y Potosí y, además, se internó en la región andina para recorrer Lluta, Socoroma, Putre, Tocrama, Lagnama, Lupica, Sacsama, Timar, Codpa, Cibitaya, Isquiña, Pachica, San Francisco de Umagata, San Santiago de Umagata, Chapiquiña y Azapa. Una primera confirmación del religioso carmelita fue que la mayor parte de la población regional vivía en la parte alto-andina y, al mismo tiempo, que era el estrato social más olvidado.

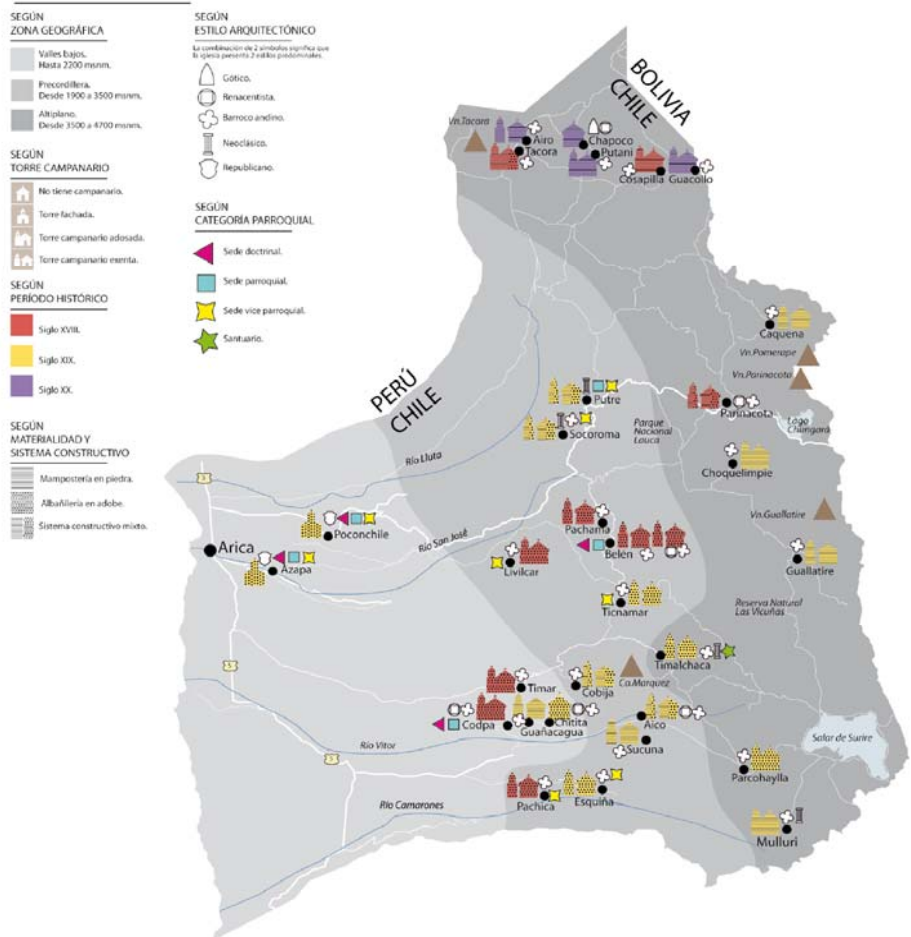

Figura 2. Mapa de la región, dividido por zonas geográficas, con la ubicación de la mayoría de las iglesias patrimoniales estudiadas y el detalle de sus estilos arquitectónicos, períodos históricos, materialidad y sistema constructivo.

En términos generales, fueron pocos los doctrineros con limitados tiempos de actuación en los entornos cordilleranos. Los valles de Lluta y Azapa aparecieron nominalmente como núcleos de control político e irradiación evangelizadora. Respecto al abandono de los templos, sin duda, sus modestas fábricas reflejaban el poco interés que tuvieron los sacerdotes por el adoctrinamiento y la promoción jerárquica personal. Mirado en su contexto, existieron también otros factores para el descuido en la asistencia y participación religiosa en el norte andino: el imperativo sísmico incorporado en la naturaleza geológica del subcontinente y la participación de la mano de obra indígena en el traslado del azogue y de la plata.

En términos conceptuales, morfológicos y de emplazamiento los incipientes conjuntos religiosos y cristianos de los Altos de Arica sintetizaron desde los albores coloniales un aspecto sencillo, tomando como referencia los aún vigentes conceptos renacentistas y barrocos, fáciles de implementar por su simpleza configurativa y estructuralmente similares a las percepciones indígenas. Los focos de irradiación arquitectónica y artística en el sur peruano colonial fueron Arequipa, Cuzco y Potosí. Los Altos de Arica mantuvieron una relación fundamental con los pueblos descritos, particularmente, desde el año 1613 con el Obispado de Arequipa.

**Las iglesias andinas de los Altos de Arica: características arquitectónicas y convivencia de estilos**

El vínculo estratégico que amparaba el propósito de la colonización fue establecido con anterioridad entre la monarquía española y el poder religioso referenciado en los papas. De esa manera, la empresa de la conquista americana fue acompañada por ambos estamentos en la escala de interrelaciones correspondiente. La obligación compartida que se desplegó en la América indígena fue clara: de una parte la corona afianzaba sus posesiones territoriales; al mismo tiempo, la iglesia aseguraba la instauración de otro credo religioso. En perspectiva, la extirpación de idolatrías fue un mecanismo visto como auspicioso que concatenó ambos propósitos. En ese sentido, desde la óptica infraestructural religiosa colonial, los mandatos provenientes tanto del Concilio de Trento como de los Concilios Limenses, fueron los mejores argumentos para la implementación de los templos y arte cristianos.

Desde los albores de la conquista de la América suroccidental, el encargo en la erección de los primeros edificios religiosos fue encomendado directamente a los caciques y poblaciones originarias. En términos prácticos, las primeras misiones de las órdenes religiosas enviadas a los entornos rurales, se enfocaron en elegir algunos puntos estratégicos desde donde abarcar grandes espacios geográficos en el que aún habitaban poblaciones indígenas dispersas: es el caso de San Jerónimo de Lluta, actual Poconchile. La implementación progresiva de las reducciones toledanas a partir del año 1572 fue un hecho concreto que marcó el fin de la diseminación poblacional en los andes sudamericanos.

Los mejores años de intercambio humano, cultural, comercial y productivo sostenidos entre Arica y Potosí durante la fase colonial, se extendieron afirmativamente hasta 1600. A partir de tal fecha, con algunas oscilaciones, prosiguieron con relativo éxito hasta 1660, empero, las actividades se prolongaron en menores proporciones hasta cerca de 1776 (año en que el Alto Perú fue escindido de su virreinato de origen). Fueron días difíciles en coincidencia con el decaimiento de la mano de obra, la abolición de la mita minera, las consecuencias de la mortífera epidemia de los años 1719-1721 y la pérdida de la ley del mineral. La ruta de la plata, sostenida en la época colonial entre el puerto de Arica y la ciudad de Potosí, significó el fortalecimiento de un circuito comercial que redundó en la articulación territorial del norte chileno y el oeste boliviano. Como no pudo haber sido de otro modo, la vigencia de tales mecanismos de intercambio ocasionó colateralmente una suma de influencias e interacciones que tuvieron en el campo de la arquitectura religiosa una contraparte de producciones materiales e inmateriales que aún permanecen como huellas y puntos de apoyo de singular valor cultural.

El dinamismo ejercido por el circuito de la plata alentó a los caciques locales en la construcción y mantención de los conjuntos religiosos, permitiendo la contratación de alarifes y artistas provenientes de latitudes cercanas que generaron una producción arquitectónica y artística prolífica. Las fábricas religiosas virreinales de los Altos de Arica muestran la convivencia y pervivencia de diversos estilos constructivos, que superan la influencia tardo-renacentista, pasando por el barroco y acogiendo los conceptos neoclásicos. Es de destacar que en el decurso histórico de las manufacturas religiosas, los simbolismos indígenas fueron permanentemente reinterpretados.

La arquitectura religiosa que empezó a materializarse en la región surandina, incluidos los Altos de Arica y sus singularidades de entorno, se afianzó en una serie de características que podemos ordenar en forma cronológica. En primer lugar, los atisbos arquitectónicos procedentes del siglo XVII no fueron pretenciosos en lo funcional y resultaron sencillos y modestos en lo que al aspecto constructivo se refiere. Se escogieron plantas uniespaciales alargadas, ábsides ochavados, muros de piedra sin cantear, cubiertas a dos aguas con soportes de par y nudillo, arcos torales de separación y techumbres de paja. En las áreas exteriores a los templos, aparecieron las cruces catequísticas, los atrios arqueados, las torres-campanario exentas, las capillas misereres y posas (Fig. 3). En los entornos circundantes a las fábricas religiosas se hicieron notar con claridad las demandas indígenas materializadas en el uso sacralizado de los espacios abiertos, formalizándose la cohabitación entre las ideologías morfológicas europeas e indígenas (Fundación Altiplano 2012).

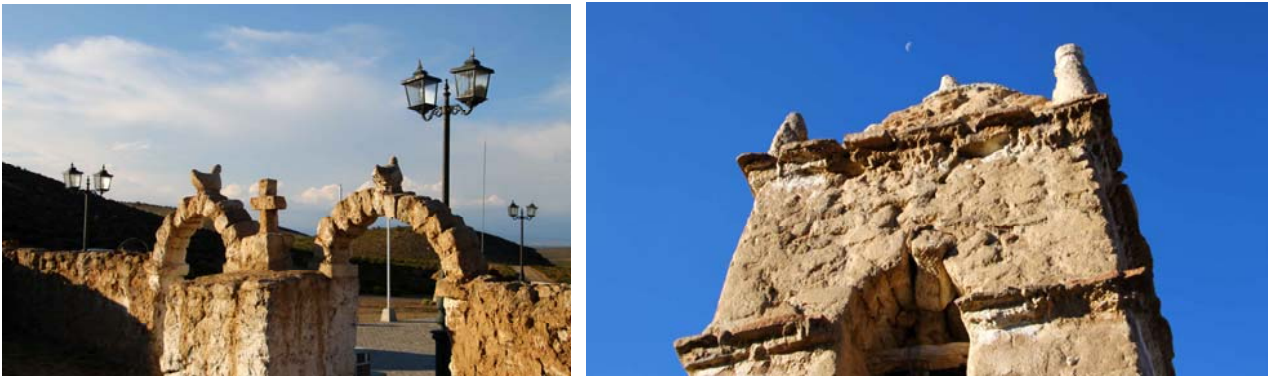

Figura 3. El atrio con arco doble y una cruz miserere o de misericordia (imagen izquierda) y la sección superior de la torre campanario adosada a la capilla de la Virgen del Rosario de Cosapilla (imagen derecha). Por encima de la torre llega a verse la Luna en cuarto creciente sobre el fondo celeste vibrante del cielo diáfano del altiplano.

A partir de la segunda mitad del siglo XVII, las exigencias programáticas en las nuevas fábricas se enfatizaron en plantas con cruz latina, muros preferencialmente canteados, cubiertas abovedadas (en algunos casos cúpulas), cruceros, capillas laterales, torres adosadas de planta cuadrada y lógica formal-estructural en la fachada (portada-retablo). Las antiguas fábricas (templos renacentistas) tuvieron que reconceptualizarse, mudando las plantas uniespaciales a la cruz latina o, en ejemplos singulares, modificando las portadas; en cualquier caso se produjo la primera convivencia de dos estilos. Finalmente, desde el último tercio del siglo XVII al segundo tercio del siglo XVIII, y con proyecciones asincrónicas hacia el primer tercio del siglo XIX, se destacó una connotación formal-estructural que resaltó lo planiforme y textilográfico de las portadas. Tal forma de pensar y hacer fue tipificada como "barroco mestizo", y su contexto espacio-temporal de vigencia se ubicó entre los últimos tercios de los siglos XVII y XVIII. La irradiación estilística se aproximó hasta Apurímac, sur del Cuzco, altiplano de Puno, Moquegua, Tacna y norte alto-andino de Chile. Los rasgos más significativos de su conformación estuvieron en la madurez de los artistas y alarifes indígenas quienes fueron capaces de conciliar las exigencias de los patrocinadores con sus destrezas creativas autóctonas, en una exigente renovación formal, funcional y volumétrica, y en la consolidación del circuito de la plata (o espacio del trajín).

**Arqueoastronomía de las iglesias andinas**

Desde antes del siglo XIX, existe un claro entrelazamiento simbólico cultural en la propia identidad de las iglesias, y que ha permanecido grabado en los templos con huellas notables. Testimonios del pasado, los elementos y símbolos ornamentales europeos conviven con los autóctonos. Como ejemplos representativos podemos mencionar: las vizcachas talladas en las columnas salomónicas de la portada de algunas iglesias, como la de Livílcar, o las decoraciones de animales y vegetales nativos presentes en la portada de Santiago Apóstol en el poblado de Belén (Fig. 4), junto a pinturas murales que muestran la imagen de un trifronte en la escena de las postrimerías de la iglesia de Parinacota.

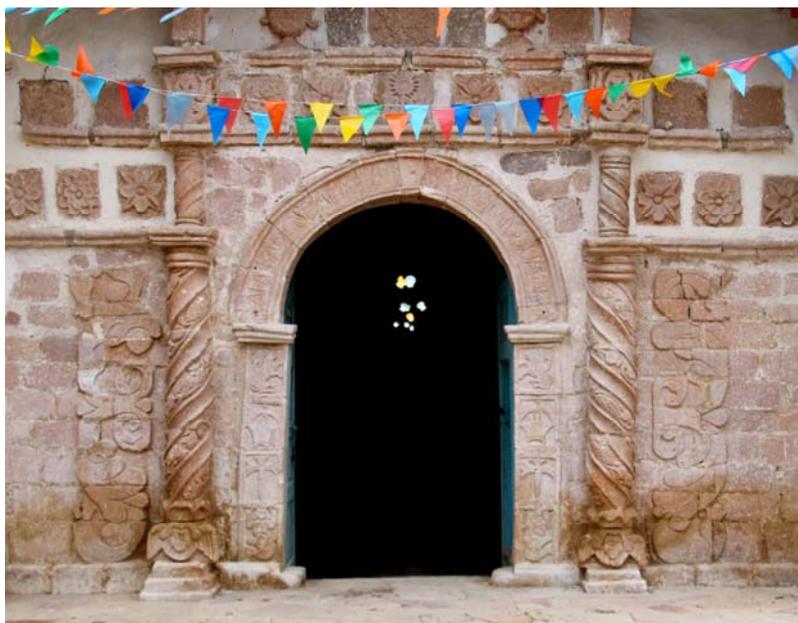

Figura 4. Varios símbolos ornamentales europeos conviven con los autóctonos. En esta imagen de la portada de la iglesia de Santiago Apóstol en Belén, dos columnas salomónicas con decoración vegetal y animal son flanqueadas por tallas de ornamentación vegetal. En las decoraciones más cercanas al pórtico se observan también los emblemas reales de Santiago: de arriba hacia abajo, el león, la corona real y la espada. Aparecen además las cabezas de ángeles alados y sirenas tenantes.

El análisis del conjunto patrimonial de iglesias andinas de la región de Arica y Parinacota es un capítulo importante en las historias del arte y de la iglesia, así como también lo es en la arquitectura y en la arqueología del paisaje andino. En esta parte nos concentraremos en el estudio de los emplazamientos de los templos en el paisaje de la región, poniendo énfasis en analizar la posibilidad de que sus orientaciones hayan sido dictadas por elementos del entorno terrestre, como volcanes, o del cielo, por ejemplo si los ejes principales de las iglesias fueron elegidos en coincidencia con los sitios del horizonte por donde sale el Sol en algún día especial del año.

El número de monumentos históricos documentados es cercano a los cuarenta y es, por lo tanto, estadísticamente significativo para un trabajo de búsqueda de correlaciones. Pensamos, además, que nuestras mediciones pueden ser representativas de la mayoría de los templos cristianos de la región, incluso cuando excedemos los límites del Chile actual. Entre otros aspectos, es nuestro interés indagar si la construcción de las iglesias de la región fue influenciada por factores tales como la presencia de la cultura y población aymara local que, en principio, fue heredera de unos patrones de culto muy diferentes a los de los colonizadores (Gisbert 1999).

En otro trabajo hemos detallado la metodología y los datos arqueoastronómicos obtenidos durante nuestra campaña de observaciones (Gangui et al. 2016). El análisis de estos datos se traduce en la Figura 5, donde se muestra el diagrama de orientación para las iglesias y capillas estudiadas.

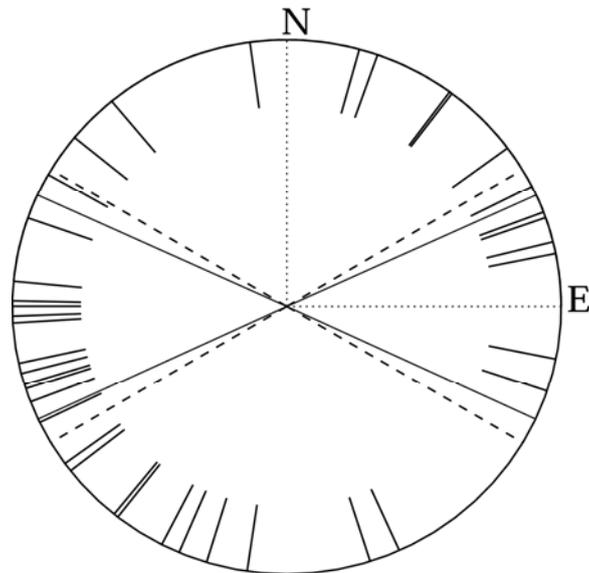

Figura 5. Diagrama de orientación para las iglesias y capillas andinas de la región de Arica y Parinacota. Las líneas diagonales del gráfico señalan los acimutes correspondientes −en el cuadrante oriental− a los valores extremos para el Sol (acimutes de 65.4° y 115.0° −líneas continuas−, equivalente a los solsticios de invierno y verano austral, respectivamente) y para la Luna (acimutes: 59.2° y 120.7° −líneas rayadas−, equivalente a la posición de los lunasticios mayores o paradas mayores de la Luna). Existe una gran diversidad de orientaciones. Sin embargo, un número significativo (un poco menos de la mitad) sigue el patrón canónico de orientación en el rango solar. Más detalles y otros análisis estadísticos en términos de declinaciones astronómicas, se hallan en el trabajo (Gangui et al. 2016).

Hemos visto que, de las 38 iglesias medidas, seis se encuentran orientadas hacia el cuadrante norte (es decir, entre acimutes 315° y 45°). Hay ocho orientadas en el cuadrante a levante (seis de ellas en el rango solar) y 16 orientadas en el cuadrante a poniente (11 de ellas en el rango solar). Finalmente, hay ocho iglesias en el cuadrante meridional (entre acimutes 135° y 225°). Entendemos que nuestra muestra es representativa de la mayoría de las iglesias de la región. En nuestros datos se distingue una orientación destacada, hacia el cuadrante occidental, con 16 iglesias que se alinean dentro de ese rango de acimutes. Entre estas, hay 11 iglesias que entran dentro del rango solar, entre los acimutes 245.0° y 294.6°.

Tengamos en cuenta, sin embargo, que en muchos de los sitios estudiados las condiciones geográficas son singulares y seguramente han jugado un papel relevante en el emplazamiento de las iglesias. Creemos que tal es el caso de la iglesia de Chitita, situada en el cañón del río Vítor y rodeada de montañas bajas, cuya orientación (con el eje alineado en 299.5° de acimut) seguramente sigue de cerca la dirección de este curso de agua. Algo similar ocurre con la iglesia de Aico, ubicada en el valle interior de la quebrada de Atco (Fig. 6). Aquí también la orientación (69.5° de acimut) parece seguir aproximadamente la traza de un río.

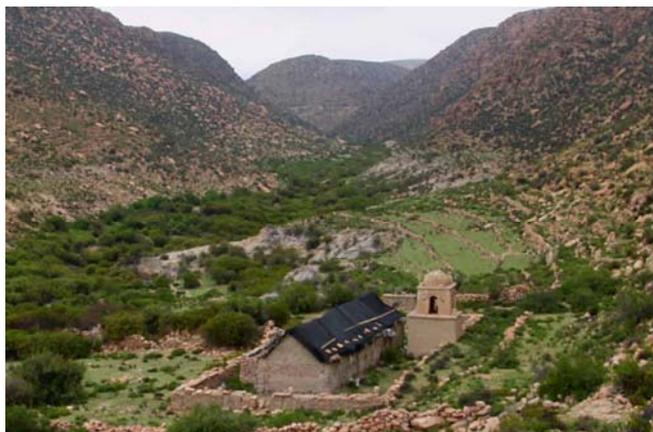

Figura 6. La iglesia San Antonio de Padua de Aico se halla emplazada en cercanías del río que recorre la quebrada de Atco, y está orientada a lo largo del valle.

Un caso diferente es el de la iglesia San Antonio de Padua, en el poblado de Sucuna, ubicada en una planicie muy cercana a la quebrada de Sacuna pero aparentemente no condicionada por elementos topográficos. En este caso, la iglesia se orienta con un acimut de 275.5°, su eje interseca el horizonte a pocos grados del lugar de la puesta del Sol en los equinoccios y su declinación calculada entra bien en el rango solar. Algo similar sucede con la capilla de Saguara, orientada también hacia el poniente (266.5° de acimut), y con la iglesia del poblado de Pachica, ubicada sobre una colina que desciende hacia la quebrada de Saguara, la que a su vez lleva a la quebrada de Camarones, pero en este caso con el eje de la construcción orientado hacia el levante (76° de acimut).

Aparte de los cauces de ríos y quebradas ya mencionados, los numerosos volcanes y picos nevados de la región pueden servir de referencia –como cerros tutelares o Apus (Reinhard 1983), incluso relacionados con el culto a los ancestros (Bouysse-Cassagne y Chacama 2012)– a la hora de decidir el emplazamiento y la orientación de los templos, por lo que conviene inspeccionar nuestros datos para verificar esa posible orientación orográfica. En el caso de la capilla Virgen del Carmen de Ungallire, al salir de su interior, en el frente, uno se encuentra con la visión imponente de los volcanes Pomerape y Parinacota; estos se ubican a solo unos pocos grados a uno y otro lado del eje principal de la construcción (Fig. 7).

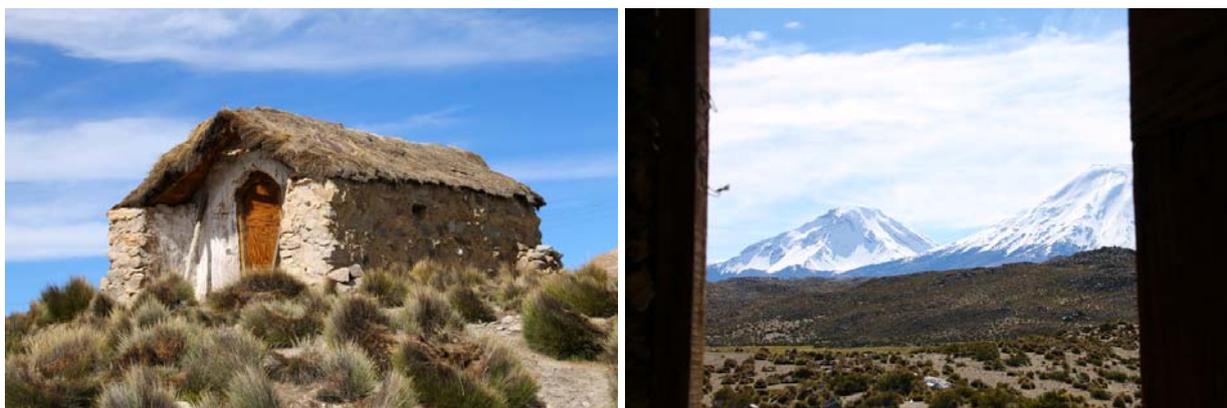

Figura 7. La capilla de Ungallire en su emplazamiento en cercanías del poblado de Parinacota (imagen izquierda) y la visión de los nevados de payachatas al salir del templo (imagen derecha), en dirección aproximadamente igual al de su eje, pero en sentido opuesto al del altar.

Se verifica algo similar con las iglesias Virgen de la Inmaculada Concepción de Guallatire y de Ancuta: de acuerdo a nuestras mediciones, el volcán Guallatiri se encuentra casi alineado con los ejes de estas iglesias pero, en ambos casos, y como sucedía con la capilla de Ungallire, la ubicación del volcán está en el frente de las construcciones (Fig. 8).

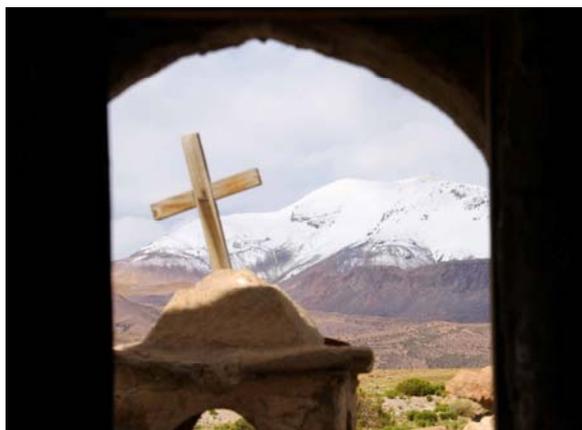

Figura 8. En la capilla de Ancuta, al salir del templo uno se encuentra frente al volcán Guallatiri.

Sin embargo, esto es muy diferente de lo que sucede con otras construcciones, por ejemplo, con la iglesia de Tacora; aunque ésta tiene al volcán del mismo nombre muy cerca de su emplazamiento, la línea de su eje difiere en unos 30° respecto de la dirección en que se halla dicha elevación prominente del paisaje. Lo mismo encontramos con la iglesia de Parinacota y su volcán homónimo, por lo que creemos que la presencia de los volcanes no parece haber sido un factor influyente al momento de orientar estos templos.

En base a este análisis, pensamos que, aunque no siempre, en muchas ocasiones (por ejemplo, en los mencionados templos de Chitita, Aico, Ungallire, Guallatire y Ancuta) las características propias –la topografía, el paisaje circundante– de cada sitio donde fueron instaladas las iglesias andinas primaron sobre las tradiciones europeas y coloniales en lo referente a la orientación de sus ejes principales, un dato que –en algunos casos, como dijimos– nos trae a la memoria más el culto aymara (Bouysse-Cassagne 1987) que las *Instrucciones* del cardenal Borromeo.[1]

**Discusión**

Pese a poseer un origen común, como respuesta evangelizadora para multitudes de trajinantes que pasaban por los nuevos poblados diseminados a lo largo de la ruta de la plata, las iglesias andinas de Arica y Parinacota muestran una cierta diversidad. Esto es aparente en varios planos, ya sea en el de sus materialidades y sistema constructivo, como en lo que respecta a sus ornamentos y arquitectura general. Por otra parte, los emplazamientos geográficos elegidos en esta extensa región y los paisajes diversos en donde estas construcciones se hayan inmersas, proveen datos adicionales que nos ayudan a pensar en las características globales de los templos cristianos.

Con la intención de sumar nuevos elementos para reflexionar sobre estas iglesias, en nuestro proyecto hemos medido sus orientaciones espaciales precisas, empleando los métodos de la arqueoastronomía. Nuestros resultados muestran que, a diferencia de lo que se encuentra en el marco de estudios llevados a cabo con iglesias antiguas europeas (González-García 2014), por mencionar un ejemplo, en las iglesias andinas patrimoniales medidas aquí no se siguió un único patrón de orientación determinante en toda la región. Sin embargo, hemos hallado que casi la mitad de las iglesias estudiadas posee una orientación que cae dentro del rango solar, con una proporción dominante en aquellas que presentan su altar hacia el poniente. Hemos también señalado algunos casos notables en los que las orientaciones de los templos parecen obedecer más a la ubicación de elementos distintivos del paisaje terrestre –volcanes como cerros tutelares– que a la salida o puesta del Sol en fechas relevantes para la advocación particular de las iglesias.

---

[1] La influencia más directa para la construcción de templos en nuestra área de estudio la encontramos en el siglo XVI, luego del Concilio de Trento, en que el Cardenal Carlos Borromeo publica sus *Instrucciones de la fábrica y del ajuar eclesiásticos* (traducidas y publicadas por Bulmaro Reyes Coria en México en 1985), con gran difusión en la época.

El presente ha sido un primer intento de caracterizar un aspecto nuevo –eventualmente astronómico– de las iglesias andinas. Sabemos que el grupo de las iglesias patrimoniales, cuyo número fue adecuado para nuestro abordaje estadístico, no agota el trabajo que se debería realizar en el futuro. La región de Arica y Parinacota está culturalmente muy emparentada con zonas del oeste boliviano y del sur del Perú, tanto en su geografía como en lo relacionado con las costumbres y religiosidad de sus pueblos. Pensamos, por ello, que el estudio de las iglesias antiguas de estas regiones cercanas puede brindarnos información adicional sobre aspectos culturales que nos permitirán comprender mejor la convivencia de costumbres en el caso de estas construcciones emblemáticas del territorio surandino.